\documentclass{ptephy}

\newcommand{\Slash}[1]{{\ooalign{\hfil$#1$\hfil\crcr\raise.167ex\hbox{/}}}}

\usepackage{graphicx}
\usepackage{wrapft}
\usepackage{lscape}

\begin{document}
\title{Hadron-Quark Crossover and Hot Neutron Stars \\
at Birth}

\author{
\name{Kota Masuda}{1,2,\ast}, 
\name{Tetsuo Hatsuda}{2,3},
\name{Tatsuyuki Takatsuka}{2}
\thanks{These authors contributed equally to this work.}
}

\address{
\affil{1}{Department of Physics, The University of Tokyo, Tokyo 113-0033, Japan}
\affil{2}{Theoretical Research Division, Nishina Center, RIKEN, Wako 351-0198, Japan}
\affil{3}{Kavli IPMU (WPI), The University of Tokyo, Chiba 277-8583, Japan}
\email{masuda@nt.phys.s.u-tokyo.ac.jp}
}

\begin{abstract}
We construct a new isentropic equation of state (EOS) at finite temperature ``{\bf CRover}'' on the basis of 
the hadron-quark crossover at high density.
By using the new EOS, we study the structure of  hot neutron stars at birth   
 with the typical lepton fraction ($Y_l=0.3-0.4$)  and the typical entropy per baryon ($\hat{S}=1-2$).
Due to the gradual appearance of quark degrees of freedom at high density, 
the temperature $T$ and the baryon density $\rho$ at the center of the hot neutron stars with the hadron-quark
crossover are found to be smaller than those without the crossover by a factor of 2 or more. 
 Typical energy release due to the contraction of a hot neutron star to a cold neutron star
  with the mass $M=1.4M_{\odot}$ is shown to be about  $0.04M_{\odot}$ with the spin-up rate about 14 \%. 
\end{abstract}

\subjectindex{Neutron stars, Nuclear matter aspects in nuclear astrophysics, Hadrons and quarks in nuclear matter, Quark matter}

\maketitle

{\bf Introduction:}
In the core-collapsed Type-II supernova explosion, the proto-neutron star (PNS) with the radius $\sim$ 100$-$200 km is formed. During the first few seconds after the core bounce, the PNS undergoes a rapid contraction and 
  evolves into either a ``hot'' neutron star (NS) with the radius $\sim$ 10$-$20 km or a black hole. 
 The hot NS at birth in quasi-hydrostatic equilibrium is composed of the supernova matter 
  with the typical lepton fraction, $Y_l=Y_e+Y_{\nu}\sim 0.3-0.4$, and 
    the typical entropy per baryon\footnote{Throughout this Letter, we put hat-symbol for 
  the thermodynamic quantities per baryon.}, $\hat{S}\sim 1-2$: They are
   caused by the neutrino trapping at the baryon density $\rho$ exceeding $10^{12} {\rm g/cm}^3$.  
  With this as an initial condition, the hot NS contracts gradually by 
  the neutrino diffusion  with the time scale of several tens of seconds and evolves to a 
 nearly  ``cold'' NS with $Y_{\nu}\simeq 0$ and $\hat{S}\simeq 0$), unless another collapse to a black hole does not take place    \cite{Prakash:2000jr,Janka:2012wk,Roberts:2012zza}.
 
 The hot neutron star provides us with various  information 
  on the properties and dynamics of high density matter  \cite{PNS:takatsuka,Roberts:2011yw}.     
 The purpose of this Letter is to study the hot neutron star at birth with degenerate neutrinos
 on the basis of 
an equation of state (EOS)  "{\bf CRover}" at finite temperature $T$ 
which is newly developed on the basis of the hadron-quark crossover picture at high baryon densities.  
 Such an EOS for ``cold'' neutron-star matter has been previously 
 studied by the present authors \cite{Masuda:2012kf,Masuda:2012ed}: 
 It was shown that a smooth crossover from the  hadronic matter to the strongly-interacting quark matter around $\rho \sim 3 \rho_0$ 
 ($\rho_0=0.17/{\rm fm}^3$ being the normal nuclear matter density) can support the 
 cold neutron star with the maximum mass greater than the observed masses of massive neutron stars,
 $M = (1.97\pm0.04) M_{\odot}$ \cite{Demorest:2010bx} and $M = (2.01\pm 0.04) M_{\odot}$ \cite{Antoniadis:2013pzd}.
 Similar conclusions for cold neutron stars have been recently
 reported by several other groups \cite{Alvarez-Castillo:2013spa,Hell:2014xva,Kojo:2014rca}.  

 In this Letter, we will generalize the idea of the hadron-quark crossover to the system
  at finite $T$ by considering the Helmholtz free energy $F(N,V,T)$ as a basic quantity
 to interpolate the hadronic matter and the quark matter. Here $N$ and $V$ are the 
 total baryon number and the total volume, respectively.
  By using our EOS for supernova matter with fixed $Y_l$ and $\hat{S}$, 
 we study not only 
 the bulk observables such as $M$-$R$ relation of the hot neutron star at birth, 
 but also  the temperature, density and  sound velocity profiles inside the star.
 We note here that  the hot neutron stars with the hadron-quark mixed phase has been studied previously, e.g. \cite{Prakash:1995uw,Nakazato:2008su,pagliara,Chen:2013tfa}. 
 Such a  mixed phase generally leads to soft EOS, so that it is rather difficult to sustain  $2M_{\odot}$ NSs.   
 On the contrary, our hadron-quark crossover approach does not suffer from the problem, since it leads to stiff EOS in the crossover region.
 
\
  
{\bf EOS for supernova matter at finite $T$ with hadrons and leptons:}
A set of finite-temperature Hartree-Fock (HF) equations is solved for
an isothermal matter composed of $n,p,e^{-},e^{+},\nu_e$ and $\bar{\nu}_e$, under 
the conditions of charge neutrality, chemical equilibrium and baryon and lepton
number conservations. It gives  the single-particle energies, chemical potentials and mixing ratios of
  nucleons and leptons, so that  the hadron+lepton EOS (HL-EOS)
 for a given set of $T$, $\rho$ and $Y_l$ is obtained.
   Then, we calculate  
 the thermodynamic quantities  such as the internal
energy $E$, the entropy $S$ and the free energy $F(=E-TS)$ (details are shown in \cite{Takatsuka:1994pm}). 
We neglect hyperons ($Y$) in the hadronic matter in this Letter, 
since they do not appear as hadronic degrees of freedom once the 
hadron-quark crossover takes place at around $\rho \simeq 3\rho_0$ as shown 
 in \cite{Masuda:2012kf,Masuda:2012ed}. 

We employ $\tilde{V}_{NN}(=\tilde{V}_{\rm RSC}+\tilde{U}_{\rm TNI})$ for 
 the effective ${NN}$ interaction in the HF calculation:
Here $\tilde{V}_{\rm RSC}$ is a two-nucleon potential
 based on the  $G$-matrix calculations with the Reid-soft-core 
potential: It depends on the total baryon density and the density difference 
between neutrons and protons. On the other hand, 
$\tilde{U}_{\rm TNI}$  is based on the  phenomenological three-nucleon potential of 
Friedman-Pandharipande type \cite{Friedman} and is written in a form of 
$\rho$-dependent two-body interaction. We introduce 
$\tilde{U}_{\rm TNI}$ to assure the correct saturation of symmetric nuclear matter
with the  nuclear incompressibility $\kappa=250$ MeV  consistent with experiments. 
This is called the 
 TNI2-EOS \cite{Takatsuka:1994pm}. The maximum mass of the NS at $T=0$ with TNI2-EOS alone
  is $1.63 M_{\odot}$.
 However, the hadron-quark crossover inside the NSs can bring this well above
  $2 M_{\odot}$ as shown in \cite{Masuda:2012kf,Masuda:2012ed}.

\

{\bf EOS for supernova matter at finite $T$ with quarks and leptons:}
 To obtain the quark+lepton  EOS (QL-EOS), we follow \cite{Masuda:2012kf,Masuda:2012ed} and 
 apply the (2+1)-flavor Nambu$-$Jona-Lasinio (NJL) model,
\begin{eqnarray}
{\mathcal L}_{\rm NJL}=
\overline{q}(i \Slash \partial-m)  q +\frac{G_{_S}}{2}\sum_{a=0}^{8}[(\overline{q}\lambda^a q)^2+(\overline{q}i\gamma_5\lambda^a q)^2] 
 -G_{_D}[\mathrm{det}\overline{q}(1+\gamma_5) q+ {\rm h.c.}]-\frac{g_{_V}}{2}(\overline{q}\gamma^{\mu}q)^2. \label{eq-1}
\end{eqnarray} 
Here the quarks $q_i$ ($i=u,d,s$) with current quark masses $m_i$ 
have three colors and  three flavors.
The term proportional to $G_{_S}$ in Eq.(\ref{eq-1}) is a $U(3)_L \times U(3)_R$ symmetric
four-fermi interaction where  $\lambda^a$ are the Gell-Mann matrices with
$\lambda^0=\sqrt{2/3}\ {\rm I}$. 
The  term proportional to $G_{_D}$ is the 
Kobayashi-Maskawa-'t Hooft (KMT)
six-Fermi interaction which breaks $U(1)_A$ symmetry explicitly. The last term with the 
positive coupling $g_{_V}$ represents a universal repulsion among different flavors.
A standard parameter set obtained from the hadron phenomenology in the vacuum
\cite{Hatsuda:1994pi} reads 
$\Lambda$ (UV cutoff)=631.4 MeV, $G_{_S}\Lambda^2$=3.67, $G_{_D}\Lambda^5$=9.29,
 $m_{u,d}$=5.5 MeV, and $m_s$ =135.7 {\rm MeV}.  As for the universal
 vector coupling, we consider the typical value, 
$g_{_V}/G_{_S}=0.5$.  The maximum mass of cold NS  with hadron-quark crossover with 
these parameters  reads 2.6$M_{\odot}$ (see Fig.19 and Table 7 of \cite{Masuda:2012ed}). 

\begin{figure}[!b]
\centering
\includegraphics*[width=9cm,keepaspectratio,clip]{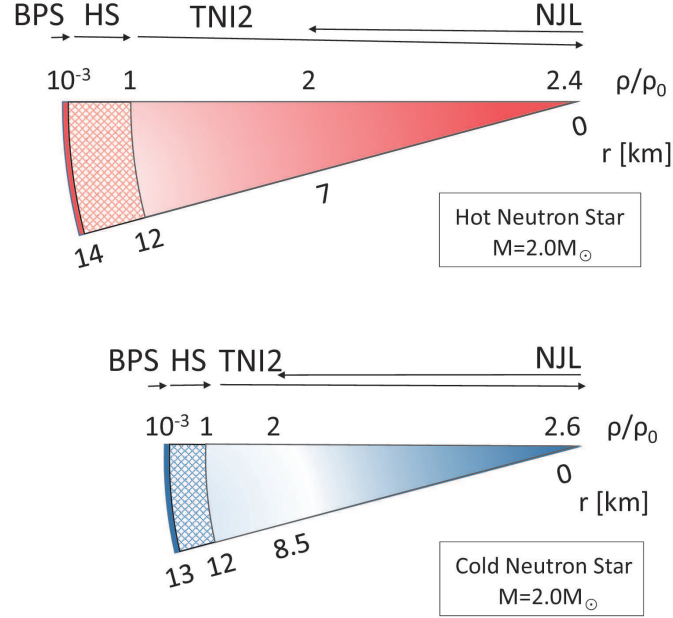}
\caption{\footnotesize{
Schematic illustrations of 
internal structure and 
relevant EOS 
in hot and cold NSs  with $M=2M_{\odot}$ obtained by  a new EOS "{\bf CRover}" with hadron-quark crossover at finite temperature.
   We take $(Y_l, \hat{S})=(0.3, 1)$ for the hot NS in this figure.
   BPS, HS, TNI2 and NJL are the EOS adopted (details are in the text).
} }
    \label{fig:fig1}
\end{figure} 

The Gibbs free-energy $\Omega$ 
is a sum of the  quark contribution  and the 
lepton contribution, $\Omega(\mu, V, T; \mu_l)=\Omega_{\rm quark}(\mu, V, T; \mu_l)
+\Omega_{\rm lepton}(\mu, V, T; \mu_l)$ with $\mu$ and $\mu_l$ being  the baryon chemical potential and the lepton chemical potential respectively. 
\footnote{
Thermal contribution from massless gluons 
above the crossover density
 is less than 4\% of the thermal part of quarks  for $T<30$ MeV, so that we do not consider gluons in this Letter.}
 The former in the mean-field approximation is 
 \begin{eqnarray} 
 & &\frac{\Omega_{\rm quark}}{V}= \nonumber \\
 & & -T\sum_i \sum_{\ell} \int \frac{d^3p}{(2\pi)^3}\mathrm{Trln}
\left(\frac{S_i^{-1}(\omega_{\ell},{\bf{p}})}{T}\right)
+G_{_S}\sum_{i}\sigma_{i}^2+4G_{_D}\sigma_{u}\sigma_{d} \sigma_{s} 
-\frac{1}{2}g_{_V}\left( \sum_i n_i \right)^2  \label{eq:Gibbs}
\end{eqnarray} 
where $n_i = \langle q^{\dagger}_iq_i \rangle$ is the quark number density in each flavor ($i=u, d, s$).
  $S_i$ is the quark propagator, which can be written as
\begin{eqnarray}
S_i^{-1}(\omega_{\ell},{\bf{p}})=\Slash p-M_i-\gamma^0 \mu_i^{\rm{eff}},
\ \ \ \mu_i^{\rm{eff}}
\equiv \mu_i-g_{_V}\sum_j n_j \label{eq:S-prop},
\end{eqnarray}
where $\omega_{\ell}=(2\ell +1)\pi T$ is the Matsubara frequency and 
 $\mu_i^{\rm{eff}}$ is the effective chemical potential. 
 The lepton part $\Omega_{\rm lepton}(\mu, V, T; \mu_l)$ corresponds to a relativistic
  and degenerate system  of electrons, neutrinos and their anti-particles. 
 The charge neutrality and chemical equilibrium among quarks and leptons are imposed.
 The  Helmholtz free-energy $F(N, V, T; Y_l)$ is obtained from the Gibbs free-energy
 $\Omega(\mu, V, T; \mu_l)$ by the Legendre transformation.

\

{\bf Hadron-Quark crossover at finite $T$:}

\begin{figure}[!b]
\centering
\includegraphics*[width=15cm,keepaspectratio,clip]{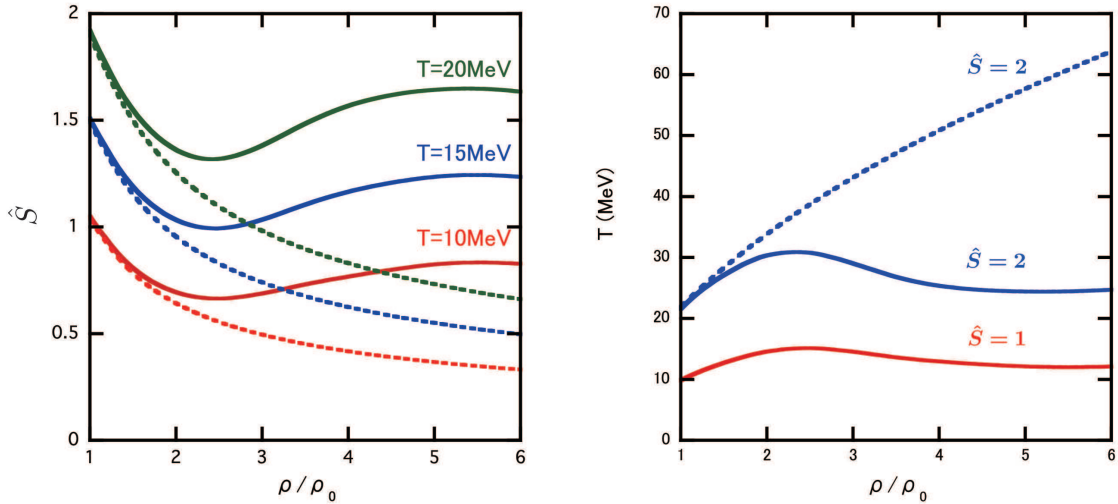}
\caption{\footnotesize{
(a) The entropy per baryon $\hat{S}$ as a function of the baryon density $\rho$ for $Y_l=0.3$ with
$T$=10 MeV (red), 15 MeV (blue) and 20 MeV (green). Solid lines and dashed lines correspond to the cases with crossover and without crossover, respectively.
(b) The temperature $T$ of the isentropic matter as a function of the baryon density $\rho$
for $Y_l=0.3$ with $\hat{S}=1$ (red) and $\hat{S}=2$ (blue).
 Solid lines and the dashed line correspond to the cases with crossover and  without crossover, respectively.
 } }
    \label{fig:fig2}
\end{figure} 

\begin{figure}[!t]
\centering
\includegraphics*[width=15cm,keepaspectratio,clip]{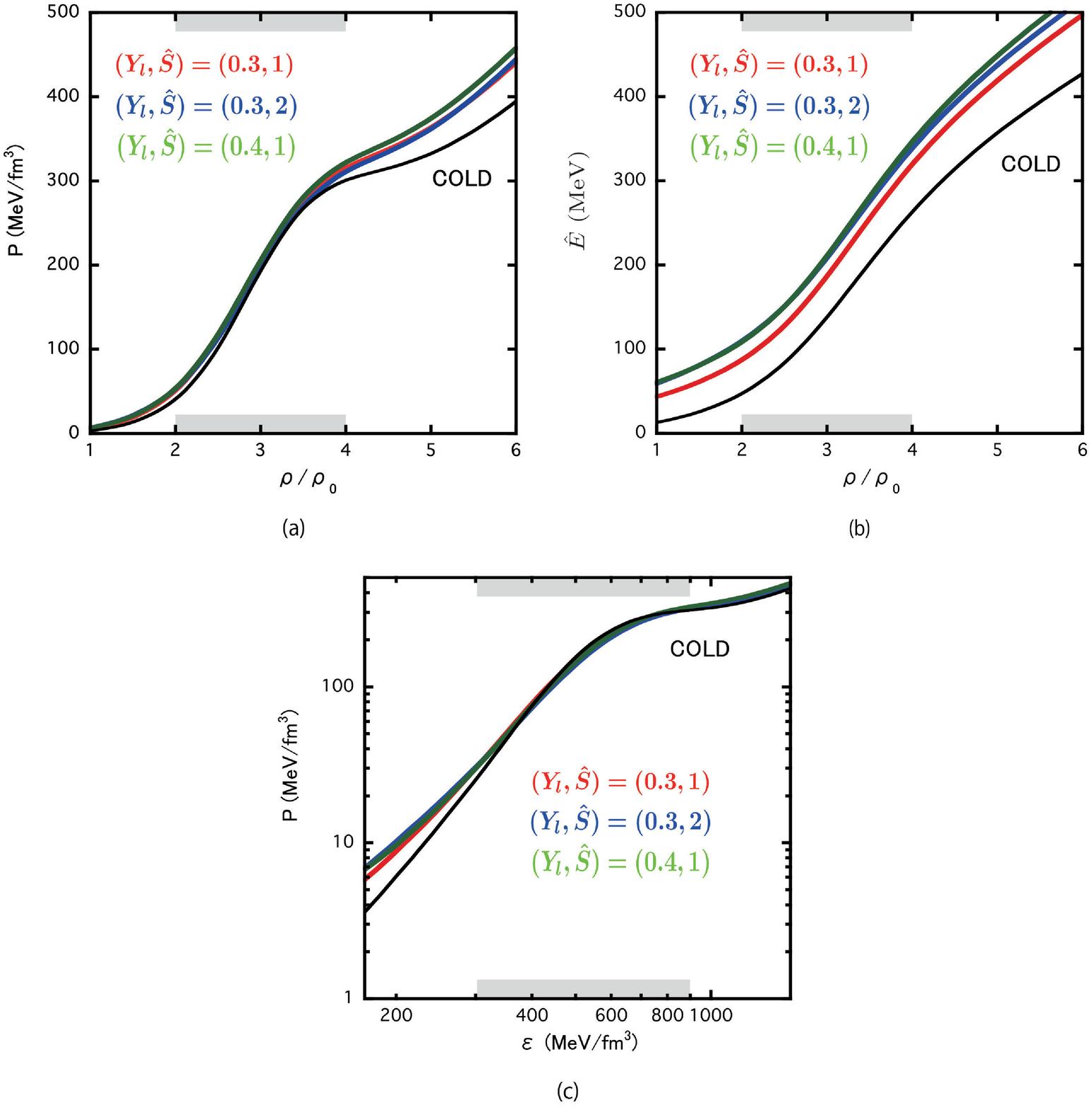}
\caption{\footnotesize{
(a)  The isentropic pressure $P$ as a function of baryon density $\rho$ for
$(Y_l, \hat{S})=$(0.3, 1), (0.3, 2) and (0.4, 1).
The black line corresponds to the EOS for cold neutron star matter.
The crossover window  is shown by the shaded area on the horizontal axis.
(b) The energy per baryon $\hat{E}$ with the same set of $Y_l$ and $\hat{S}$ as (a).
(c) $P$ as a function of $\varepsilon$.
 } }
    \label{fig:fig3}
\end{figure}  

 \begin{figure}[!h]
\centering
\includegraphics*[width=15cm,keepaspectratio,clip]{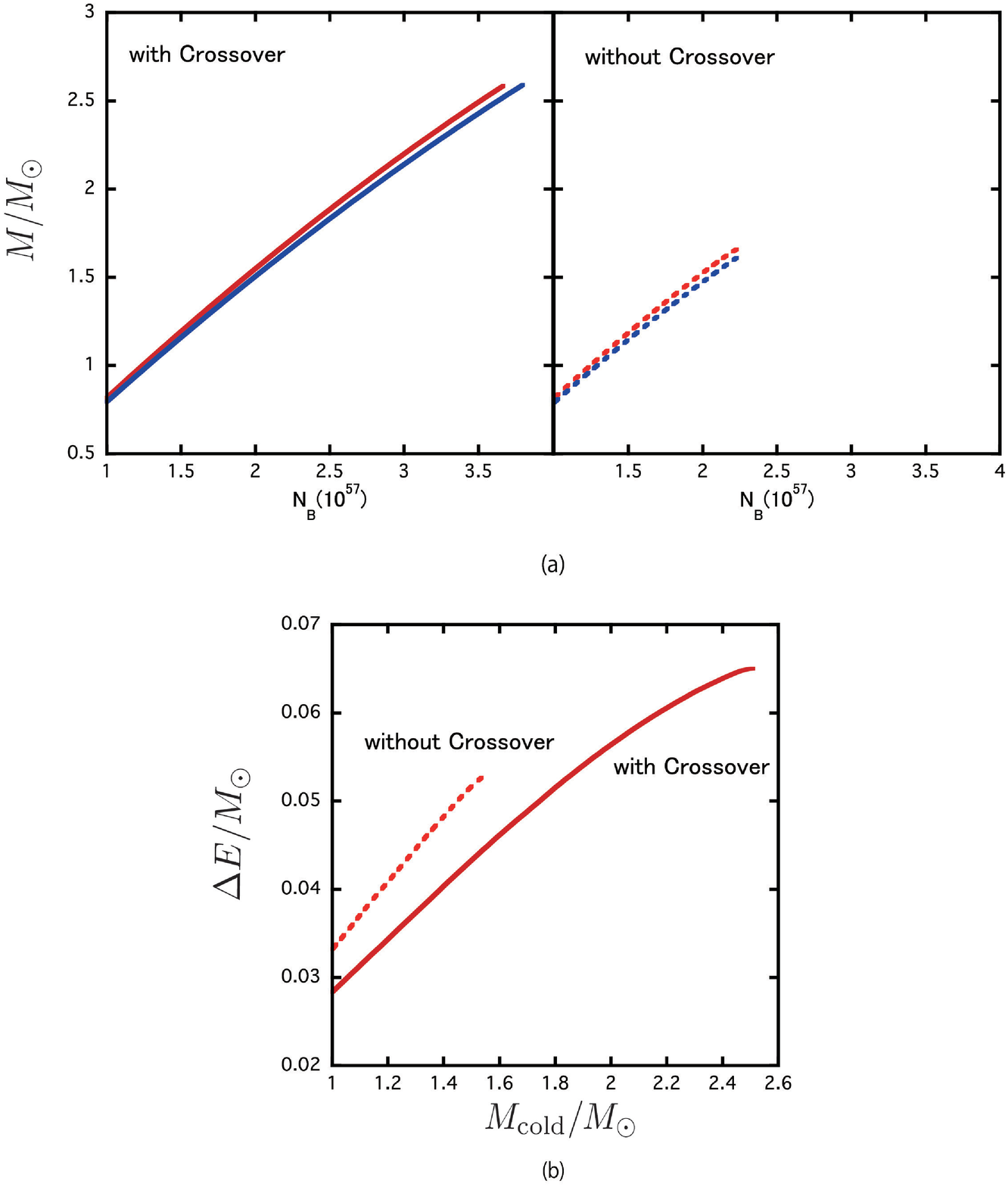}
\caption{\footnotesize{
(a) The neutron star 
mass $M$ as a function of the total baryon number $N_B$.
Red (blue) curves correspond to the hot (cold) neutron stars with crossover.
The dotted lines correspond
to the case without crossover.
(b) The energy release $\Delta E$ as a function of the cold neutron star mass $M_{\rm cold}$.
 $(Y_l, \hat{S})=(0.3, 1)$ is adopted.
 } }
    \label{fig:fig4}
\end{figure}

Let us now generalize the idea of the hadron-quark crossover introduced in our previous works
\cite{Masuda:2012kf,Masuda:2012ed} to the system at finite $T$.
We consider the following interpolation of the Helmholtz free-energy per baryon $\hat{F}$
between  the hadron+lepton free-energy per baryon ($\hat{F}_{\rm HL}$) and the quark+lepton  free-energy 
per baryon ($\hat{F}_{\rm QL}$),
\begin{eqnarray}
\hat{F}(\rho,T; Y_l)=\hat{F}_{\rm HL}(\rho,T; Y_l) w_-(\rho,T)+\hat{F}_{\rm QL}(\rho,T; Y_l)  w_+ (\rho,T).
\label{eq:new-HQ-EOS}
\end{eqnarray}
Here $w_{-}$ and $w_+=1-w_-$ are the  weight functions.
 The typical temperature we consider is  30 MeV or less.  This  
 is considerably  smaller than 
 the thermal dissociation temperature ($\sim$ 200 MeV) of  hadrons into quarks and gluons 
  \cite{Asakawa-Hatsuda},
  so that we assume  $w_{-}$ being $T$-independent and takes the same form as given in  
\cite{Masuda:2012kf,Masuda:2012ed}:
\begin{eqnarray}
w_{\pm}(\rho,T) \rightarrow w_{\pm}(\rho) 
= \frac{1}{2} \left( 1\pm \mathrm{tanh}\left( \frac{\rho-\bar{\rho}}{\Gamma} \right) \right),
\label{eq:HQ-EOS}
\end{eqnarray}
where  $\bar{\rho}$ and $\Gamma$ are
 the phenomenological interpolation parameters which 
 characterize the crossover density  and the width of the crossover window, respectively.
We take $(\bar{\rho},\Gamma)=(3\rho_0,\rho_0)$ as typical values to 
account for the 2$M_{\odot}$ neutron stars at $T=0$  \cite{Masuda:2012kf,Masuda:2012ed}.
Note that hyperons do not appear in the hadronic phase with such a  relatively low  crossover density.

From  Eq.(\ref{eq:new-HQ-EOS}) with Eq.(\ref{eq:HQ-EOS}), 
other bulk quantities, the entropy per baryon ($\hat{S}$) and the 
energy per baryon ($\hat{E}$), are derived by using thermodynamic relations,
$\hat{S}=-\partial \hat{F}/\partial T$ and $\hat{E}=\hat{F}+T\hat{S}$, as
\begin{eqnarray}
\hat{S}(\rho,T; Y_l)&=&\hat{S}_{\rm HL}(\rho,T; Y_l) w_-(\rho)+\hat{S}_{\rm QL}(\rho,T; Y_l)  w_+ (\rho), 
\label{entropy} \\
\hat{E}(\rho,T; Y_l)&=&\hat{E}_{\rm HL}(\rho,T; Y_l)w_-(\rho)+ \hat{E}_{\rm QL}(\rho,T; Y_l) w_+ (\rho).
\label{energy density} 
\end{eqnarray}

The thermodynamic quantities for isothermal matter with $\rho$, $T$ and $Y_l$ can be
 converted into those for isentropic matter by means of the $T$-$\rho$ relationship
  constrained by the constant entropy per baryon, $\hat{S}=$const. Then 
  the isentropic pressure is obtained as
 \begin{eqnarray}
 P(\rho,T(\rho);Y_l, \hat{S})= - \left. \frac{\partial E}{\partial V} \right|_{S,N}
 =  \rho^2 \left. \frac{\partial \hat{E}}{\partial \rho} \right|_{\hat{S},N} .
 \end{eqnarray}
  
\
 
{\bf EOS for supernova matter below $\rho_0$:}

 Below the normal nuclear matter density $\rho_0$, we use the thermal
 EOS which  consists of an ensemble of nuclei
 and interacting nucleons in nuclear statistical equilibrium by Hempel and Schaffner-Bielich  \cite{Hempel:2009mc}.
 We call this as HS EOS in the following.
 Various EOSs below $\rho_0$ do not show quantitative difference 
 as discussed in \cite{Buyukcizmeci:2012it}. 
  Once the baryon density becomes smaller than the neutron drip density $10^{-3} \rho_0$, 
 the temperature becomes smaller than 0.1 MeV, so that 
 we switch to  the  standard BPS EOS  \cite{Baym:1971pw}.
  In Fig.\ref{fig:fig1}, 
  the internal structures of hot and cold NSs with our isentropic EOS are illustrated. 
\

{\bf EOS for supernova matter with hadron-quark crossover:}
To carry out the conversion from the isothermal matter with fixed $T$ and $Y_l$ to
 the isentropic matter with fixed $\hat{S}$ and $Y_l$, we consider
$\hat{S}(\rho,T,Y_l)$ as a function of $\rho/\rho_0$.
In  Fig.\ref{fig:fig2} (a), we show $\hat{S}$ for $T$=10, 15, 20 MeV and $Y_l=0.3$
 with crossover (solid lines) and without crossover (dashed lines). One finds that
  the entropy per baryon turns out to increase at high densities, once 
  the quark degrees of freedom start to appear.  
 Fig.\ref{fig:fig2} (b) shows the temperature as a function 
 of $\rho/\rho_0$ under isentropic conditions for $\hat{S}=1, 2$ and $Y_l=0.3$
  with crossover (solid lines) and for $\hat{S}=2$ and $Y_l=0.3$ without crossover (the dashed line).
 Due to the appearance of the colored degrees of freedom, 
  $T$ is suppressed above the crossover density.

  In Fig.\ref{fig:fig3}(a), (b) and (c), we show the isentropic pressure $P(\rho,T(\rho),Y_l)$ and
 the energy per baryon  $\hat{E}(\rho,T(\rho),Y_l)$, and the $P-\varepsilon$ relation, respectively,
 for the three  characteristic sets, $(Y_l,\hat{S})=(0.3, 1), (0.3, 2)$ and $(0.4, 1)$.
 Here  $\varepsilon \equiv \rho \hat{E}$ is the energy density.
     For comparison, the EOS of cold neutron star matter ($T=0$ without neutrino degeneracy)
 is also shown by the black solid lines. 
 We call this new EOS with hadron-quark crossover as ``{\bf CRover}''
  and will post the numerical tables for different combinations of $\hat{S}$ and $Y_l$
 in the EOS database (EOSDB) \cite{Ishizuka:2014jsa}.

\
 
{\bf Structure of hot neutron stars:}
We now solve the Tolman-Oppenheimer-Volkov (TOV) equation 
to obtain the structure of hot neutron star at birth;
\begin{eqnarray}
\frac{dP}{dr}&=&-\frac{G}{r^2}\left(M(r)+4\pi Pr^3\right)\left(\varepsilon +P\right)\left(1-2GM(r)/r\right)^{-1} , \nonumber \\
M(r)&=&\int^r_0 4\pi r'^2 \varepsilon(r') dr' .
\end{eqnarray}
Here we have assumed spherically symmetric stars with $r$ being the 
 radial distance from the center. In the following, we will consider the hot neutron stars 
 with typical values,  $(Y_l,\hat{S})=(0.3, 1)$.

 In Fig. \ref{fig:fig4} (a), we show the relation between the gravitational mass $M$ and the 
 baryon number $N_B$ for hot and cold neutron stars with and without crossover.
 First of all, the crossover leads to heavier neutron stars for given $N_B$ due to the 
 stiffening of the EOS.  Furthermore, hot neutron stars have larger mass than the 
 cold ones for given $N_B$. 
 \footnote{
 In the approach based on the hadron-quark mixed phase, 
 the softening of the EOS due to the mixed phase is tamed by the finite temperature effect, 
 so that the maximum value of $N_B$ for hot NS becomes larger than that of the cold NS. 
 This is why there are hot NSs which eventually collapse into black holes  \cite{Prakash:1995uw,pagliara}.  
 Similar observation in the case of the pion condensation has been reported ealier by one of the present authors \cite{PNS:takatsuka}. 
 On the other hand, in our hadron-quark crossover picture, 
 the hot EOS and cold EOS are similar except for the low density region,   
 so that  the delayed collapse to the black hole does not take place. 
   }
 Typical energy release for $M_{\rm cold}=1.4 M_{\odot}$ 
  due to the contraction 
  reads  $\Delta E =M_{\rm hot} - M_{\rm cold} \sim 0.04M_{\odot}$. 
  In Fig. \ref{fig:fig4} (b), we show $\Delta E$ in the unit of $M_{\odot}$ 
  as a function of the mass of  cold neutron star, $M_{\rm cold}$.  
 
\begin{figure}[!h]
\centering
\includegraphics*[width=15cm,keepaspectratio,clip]{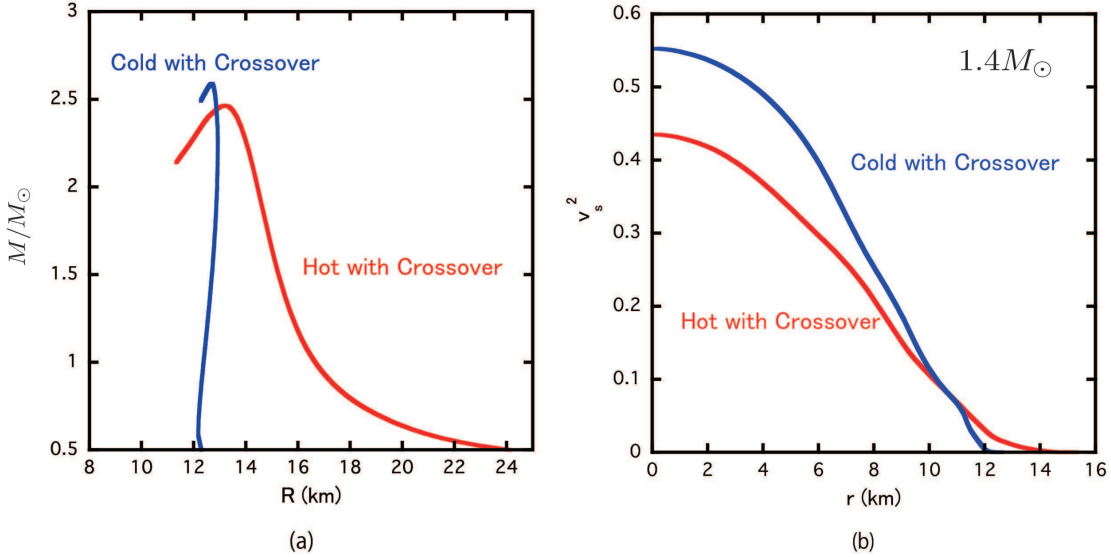}
\caption{\footnotesize{
(a) Mass-radius relationship for $(Y_l, \hat{S})=(0.3, 1)$.
Red: hot neutron stars with crossover. 
Blue: cold neutron stars with crossover. 
(b) The sound velocity squared $v_s^2$ as a function of the distance from the center 
of $1.4M_{\odot}$ neutron star.  Colors on each line are the same as in (a).
 } }
    \label{fig:fig5}
\end{figure}

 In Fig. \ref{fig:fig5} (a), the mass-radius $(M$-$R)$ relations of the hot and cold neutron stars
  with hadron-quark crossover are shown. Although the  maximum masses
 are similar between hot and cold neutron stars, the stiffening of the EOS at finite $T$ at low density
 (see Fig. \ref{fig:fig3} (c)) 
  makes the hot neutron star bigger in size, especially for neutron stars with small $M$.
 In Fig.\ref{fig:fig5} (b),
   we plot the local sound velocity squared  $v_s^2(r)$ 
 as a function of $r$ for $M=1.4M_{\odot}$ obtained by
\begin{eqnarray}
v_s^2 (\rho;Y_l, \hat{S})= \left. \frac{\partial P}{\partial \varepsilon} \right|_{Y_l,\hat{S}}
=  \left. \frac{d P(\rho,T(\rho); Y_l, \hat{S})/d\rho}{d \varepsilon(\rho,T(\rho); Y_l, \hat{S})/d\rho} \right|_{Y_l,\hat{S}},
\end{eqnarray} 
 with $\rho(r)$ obtained by the TOV equation. 
 The sound velocity for cold neutron star is larger (smaller) at higher (lower)  density
 than that of the hot neutron star.

\begin{figure}[!h]
\centering
\includegraphics*[width=15cm,keepaspectratio,clip]{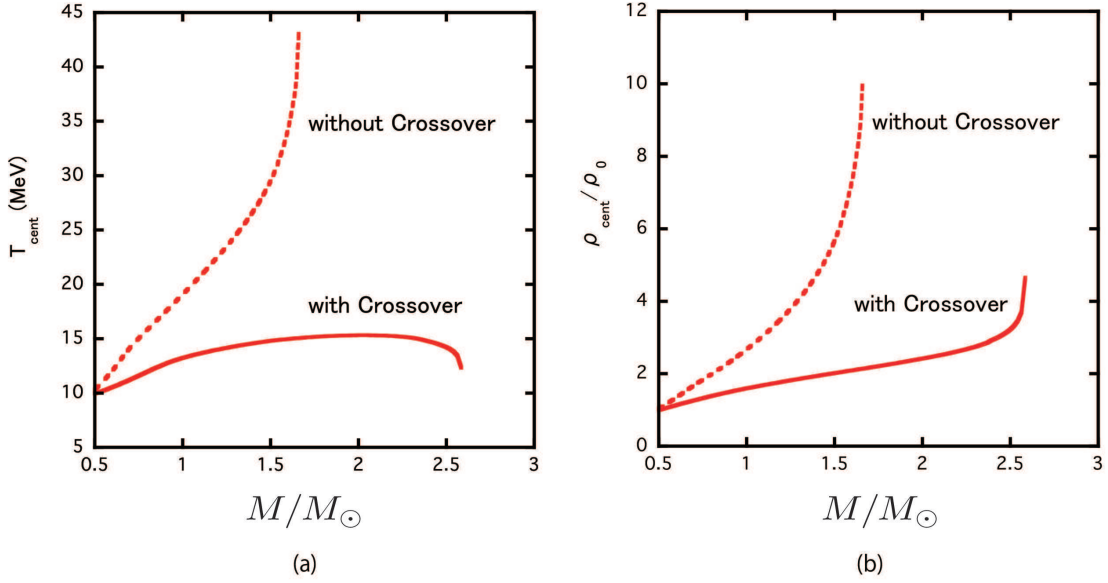}
\caption{\footnotesize{
(a) The central temperature $T_{\rm cent}$ as a function of $M$.
(b) The central density $\rho_{\rm cent}$ as a function of the neutron star mass $M$.
The solid (dashed) lines correspond to the EOS with (without) crossover, and 
 $(Y_l, \hat{S})=(0.3, 1)$ is taken.
 }}
    \label{fig:fig6}
\end{figure}

In Fig. \ref{fig:fig6}(a) and (b),
 the central temperature $T_{\rm cent}$ and  the central density $\rho_{\rm cent}$. 
   of the hot neutron stars with and without the crossover 
  are plotted as a function of $M$.    
   The temperature decreases due to the appearance of the quark degrees of freedom   
   (see Fig.\ref{fig:fig2}(b)), so that 
  the central temperature becomes significantly smaller with crossover as shown in Fig. \ref{fig:fig6}(a).
  The  central density of the star becomes significantly smaller with crossover as shown Fig. \ref{fig:fig6}(b)
  due to the fact that  the  EOS is stiffer with crossover.

\begin{figure}[!h]
\centering
\includegraphics*[width=15cm,keepaspectratio,clip]{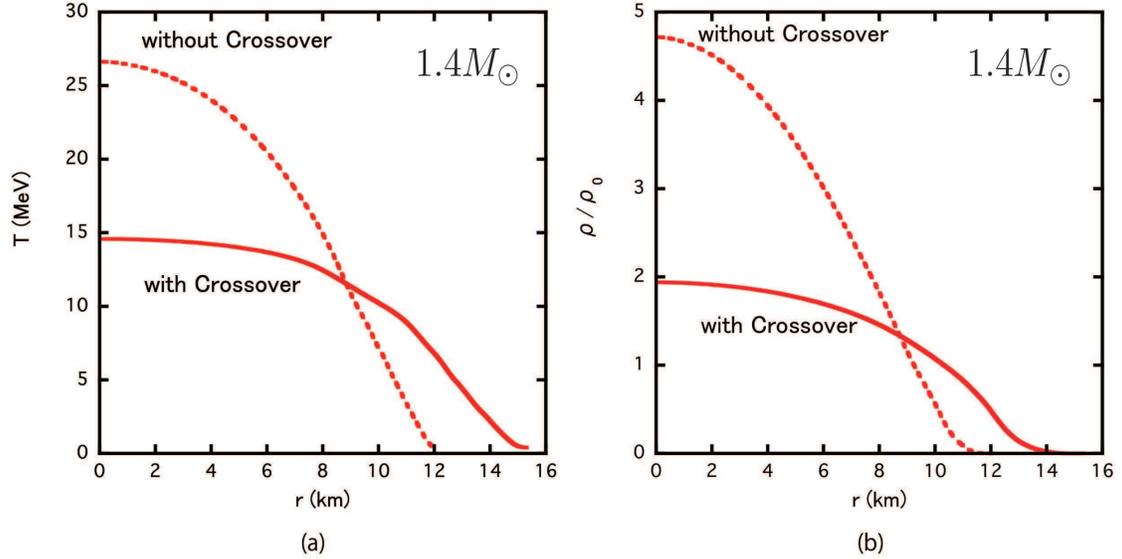}
\caption{\footnotesize{
(a) The temperature profiles of the hot neutron star with $M=1.4M_{\odot}$ and $(Y_l, \hat{S})=(0.3, 1)$.
Solid (dashed) lines correspond to the EOS with crossover and without crossover respectively.
(b) The density profiles of the same neutron star as the case (a).
 } }
    \label{fig:fig7}
\end{figure}

To compare the internal structure of the hot neutron star
  with and without the crossover in terms of the  temperature and density, 
  we plot  $T(r)$ and $\rho(r)$ 
 in Fig. \ref{fig:fig7} (a) and (b), respectively. Here we consider the hot neutron star with 
 a canonical mass $M=1.4M_{\odot}$. Under the presence of the hadron-quark crossover,
 the EOS is stiffer and hence the central density is smaller, so 
 that the temperature and density profiles of 
the star are more uniform as compared to those without the crossover.

\

{\bf Summary and concluding remarks:}
In this Letter, we have discussed the properties of hot neutron stars at birth on the basis of a new 
EOS ``{\bf CRover}'' for supernova matter with hadron-quark crossover. 
Such a crossover leads to the EOS stiff enough 
to sustain $2M_{\odot}$ neutron stars. 
 A noticeable point is that the crossover plays important roles not only to generate the stiff EOS but also 
  to lower the internal temperature of hot neutron stars.  Such suppression of temperature
   originates from a combined effect of the isentropy nature of the supernova matter and larger  
 entropy for given temperature due to the quark degrees of freedom.
 Given baryon number, hot neutron stars have larger radius and larger gravitational mass caused by
 the high lepton fraction and the thermal effect. This suggests that, during the contraction
  from hot to cold stars, gravitational energy is released and simultaneously the spin-up takes place.
 In the present study, the released energy is about 0.04 $M_{\odot}$ and the spin-up rate
 is about 14 \% (assuming the conservation of angular momentum) 
 for  $M_{\rm cold}=1.4M_{\odot}$ of evolved cold neutron stars.
 It is worth noting here that the maximum baryon number $N_{\rm max}$ is smaller for the 
  hot neutron star if there is a crossover. This implies that there could by massive neutron stars
  which can be reached only by the mass accretion after the birth.
   In the present study, we have used temperature-independent  interpolation function to construct 
  the EOS with crossover.  Although this is justified for the supernova matter with temperature
  less than 30 MeV, further investigations would be necessary to have the EOS applicable to
  wider range of temperature in supernovae and neutron star mergers.

\

{\bf Acknowledgment}

We thank Gordon Baym for helpful discussions in the early stage of this work.
KM thanks Mark Alford for the kind hospitality and discussions in Washington Univ. in St. Louis
where part of this work was carried out under the support of ALPS Program, Univ. of Tokyo.
KM is supported by JSPS Research Fellowship for Young Scientists.
 TH and TT were partially supported by JSPS Grant-in-Aid for Scientific Research, No.25287066.
 This work was partially supported by RIKEN iTHES Project.


\vspace{20pt}



\end{document}